\documentclass[10pt,a4paper,notitlepage,onesided]{article}   % druh dokumentu
\usepackage[czech]{babel}      						% načtení balíku maker, čeština
\usepackage[utf8]{inputenc}							% kódování textového editoru
\usepackage{multirow}
\usepackage{color}
\usepackage{graphicx}								% vkládání obrázku v png
\usepackage{epstopdf}
\usepackage{xcolor}
\usepackage{amsmath, bm}								% balíček pro pokročilou matem. sazbu
\usepackage{amssymb}
\usepackage{amsthm}	
\usepackage{bm} 						
\usepackage{comment}
\usepackage{array}
\usepackage{wasysym}								% umožní použití znaku průměr
\usepackage{caption}
\usepackage{float}
\usepackage{multicol}
\usepackage{subcaption}
\usepackage{cite}				% pomlčka mezi více citacemi, pokud se použije \cite{art1, art2, art3}
\begin{document}

\renewcommand{\figurename}{Fig.}			
\renewcommand{\tablename}{Tab.}
\renewcommand{\refname}{References}

\noindent \textbf{\huge Single-shot spatial coherence of a plasma based soft X-ray laser}

\vspace{36pt}

\noindent M. Albrecht$^{1,2}$, O. Hort$^{1}$, M. Kozlová$^{1,3}$, M. Krůs$^{3}$ and J. Nejdl$^{1,2*}$
\\
\newline 
\begin{scriptsize} 
$^1$ ELI Beamlines Facility, The Extreme Light Infrastructure ERIC, Za Radnicí 835, 
252 41 Dolní Břežany, Czech Republic \\ 
$^2$ Czech Technical University in Prague, Faculty of Nuclear Sciences and Physical Engineering, 
Brehova 7, 115 19 Prague 1, Czech Republic\\
$^3$ Institute of Plasma Physics, Czech Academy of Science, Za Slovankou 3, 182 00 Prague 8, Czech Republic\\

*Corresponding author: jaroslav.nejdl@eli-beams.eu\\
\newline
% Received XX Month XXXX; revised XX Month, XXXX; accepted XX Month XXXX; posted XX Month XXXX (Doc. ID XXXXX); 
% published XX Month XXXX
\end{scriptsize}
 
\vspace{12pt}
\hrule
\vspace{12pt}

\section*{Abstract}

\textbf{
Many applications of short-wavelength radiation impose strong requirements on the coherence 
properties of the source. However, the measurement of such properties poses a challenge, 
mainly due to the lack of high-quality optics and source fluctuations that often violate 
assumptions necessary for multi-shot or cumulative techniques. In this article, we 
present a new method of single-shot spatial coherence measurement adapted to the soft 
X-ray spectral range. Our method is based on a far-field diffraction pattern from a 
binary transmission mask consisting of a non-redundant array of simple apertures. 
Unlike all currently available methods, our technique allows measuring radiation 
field with an arbitrary spatial coherence function without any prior assumption on intensity
distribution or the model of the degree of spatial coherence. We experimentally verified the 
technique by retrieving the spatial coherence functions of individual 
shots of laser-driven Zn plasma soft X-ray laser with one- and two-dimensional 
masks. The experimental results revealed nontrivial illumination pattern 
and strong asymmetry of the spatial coherence function, which clearly calls for 
abandoning the often used models that assume rotational invariance of the coherence 
function, such as the popular Gaussian-Schell beam model.}

\vspace{12pt}
\hrule

% \section{Introduction}
\section*{}

The main characteristics of coherent waves are their deterministic behavior during propagation 
and hence a relatively simple mathematical description. The coherence of electromagnetic 
radiation is, therefore, one of its crucial characteristics that has a significant impact on various
applications based on imaging, holography, interferometry, diffraction, etc. 
The coherent properties are commonly described by the mutual coherence function, 
which reflects the statistical properties of light. Frequently, under an assumption 
of cross-spectrally pure light \cite{mandel}, this function is factorized, and temporal 
(longitudinal) and spatial (transverse) coherence can be defined. The spatial 
coherence of the field indicates its focusability and applicability, for example, in
wavefront-division interferometry \cite{gartside_2010}.

X-ray sources occurring in nature or the ones constructed by humankind possess 
significant beam fluctuations due to various nonlinear processes involved in 
their operation. In combination with lack of high-quality X-ray optics, the 
characterization of the spatial coherence of X-ray fields presents a crucial challenge
that has not been satisfactorily addressed yet. Up to now, several methods have been 
reported for investigating the spatial coherence through diffraction measurements using 
a binary mask in the soft X-ray spectral region. The traditional Young's double-slit 
experiment \cite{young_1804} that relies on field stationarity, since multiple points 
of coherence function need to be measured by a varying the distance between apertures, 
has been directly adopted for soft X-rays \cite{liu, vartanyants, singer_1, singer_2}. 
An alternative technique for evaluating the complete coherence function involves comparing 
the experimental diffraction pattern obtained from a single diffraction pattern of a binary 
transmission mask, which contains a rich content of spatial frequencies \cite{nugent}, 
with the calculated diffraction pattern of a fully coherent field. This method has been 
employed in measurements of a soft X-ray laser \cite{trebes}. However, both methods require 
knowledge of intensity distribution, which is often approximated by homogeneous illumination 
of the mask.

A more advanced method of the spatial coherence measurement is based on a specially designed 
non-redundant mask containing an array of N identical apertures \cite{gonzales, mejia}, 
which allows measuring the coherence function at $\binom{N}{2}$ points from a single 
diffraction pattern. However, this method succeeds only if the intensity distribution on 
the apertures is known or if there is an assumption of rotational invariance for both the intensity 
and coherence function \cite{j_duarte}. It is very challenging at the same time to measure 
the intensity distribution on the scattering mask and the diffraction pattern in the soft X-ray 
spectral region. As a result, the intensity measurement is typically replaced by a simplified 
model that may depart from reality.

In this article, we present, for the first time, a single-shot method for measuring spatial 
coherence without requiring prior knowledge of the intensity profile or making any 
assumptions about the coherence function itself. We experimentally demonstrate the method  
by measuring the coherence of individual shots from plasma based soft X-ray laser.

Our method for retrieving the coherence function is based on analyzing the far-field 
diffraction pattern produced by a known non-redundant array (NRA) of apertures. 
According to the theory of optical field coherence \cite{mandel}, the 
coherence function can be expressed as a sum of mutually incoherent modes, which can be 
independently propagated within a given space.
To retrieve the coherence function, we employ an iterative algorithm similar to the one 
used in coherent diffraction imaging \cite{rydberg}. This iterative process allows us 
to obtain a self-consistent representation of the coherence function that satisfies 
the constraints in both the NRA mask plane and the detector plane where the 
intensity diffraction pattern is measured (refer to the "Methods" section for more details).
%Through numerical propagation of the modes between these two planes and the application of constraints in each plane, we ensure convergence towards a set of modes that form the coherence function. 
Ultimately, this determines the degree of spatial coherence and the resulting intensity 
pattern in the mask plane.

\begin{figure}[H]

\centering  
\includegraphics[scale=1.5]
  {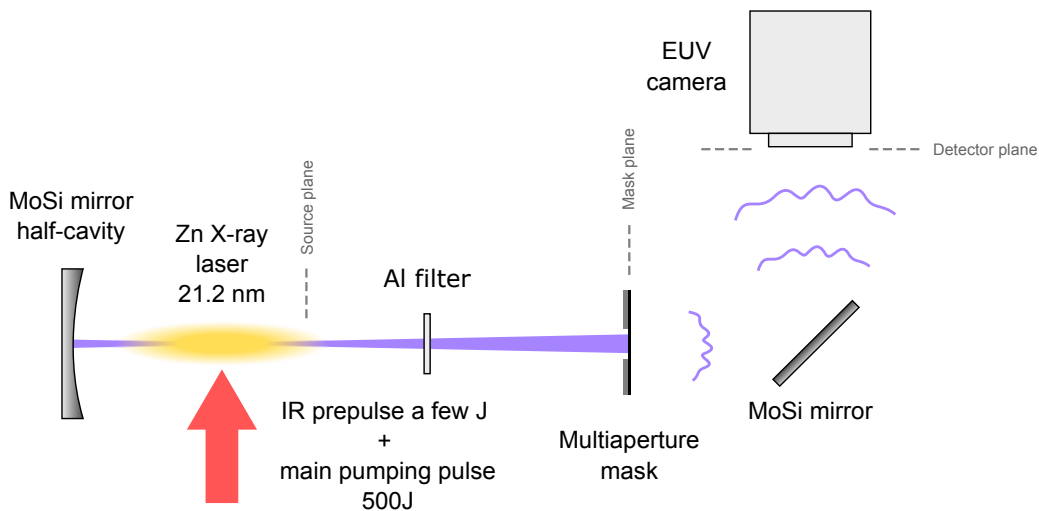}  
\caption{Experimental setup of the single-shot method for measuring the spatial coherence 
of Zn soft X-ray laser (21.2 nm).}
\label{experimental_setup_Zn_XRL}

\end{figure}

We have verified our method of single shot spatial coherence measurement using 
one-dimensional and two-dimensional masks to characterize the beam of a Ne-like 
Zn plasma soft X-ray laser (SXRL) emitting at 21.2 nm \cite{b_rus}.
The SXRL source was operated in the double pass amplification by reflecting 
the laser radiation using a MoSi multilayer mirror placed near one end of a 3 cm 
long plasma column at normal incidence. The schematics of the 
experimental setup is depicted in Fig. \ref{experimental_setup_Zn_XRL}.

The NRA consisted of slits (for the 1D measurement) or circular apertures 
(for the 2D measurement) was positioned 1.5 m downstream from the SXRL source. To eliminate 
the driving laser radiation and reflect the radiation within a narrow bandwidth 
centered at the SXRL lasing line, a 45-degree MoSi multilayer mirror with a thin 
Al filter was employed. The diffraction pattern was recorded using a back-illuminated 
CCD camera placed 6 meters behind the NRA.

\begin{figure}[H]

\centering
\includegraphics[scale=0.7]
    {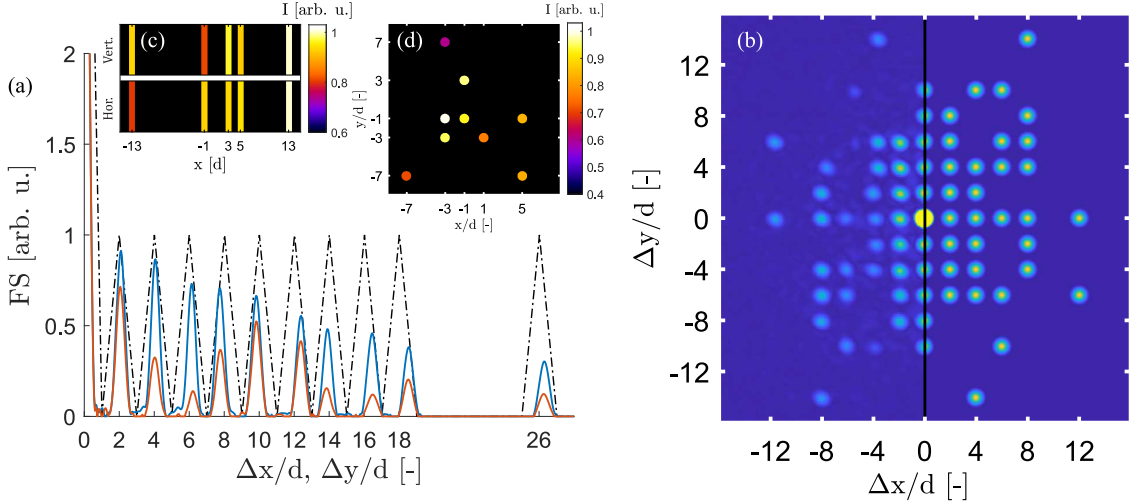}	
    
\caption{
(a): Autocorrelation function of the 1D mask of slits (dashed dotted black line) 
and Fourier spectrum of the measured diffraction pattern (FS) of the measurement 
performed in the horizontal (blue line) and the vertical orientation (red line). 
(b): Autocorrelation function of the 2D mask of circular apertures (right part of the image) 
and Fourier spectrum of the measured diffraction pattern (left part of the image). 
(c): 1D mask of slits with retrieved intensity in vertical and horizontal direction. 
(d): 2D mask of circular apertures with retrieved intensity. Both 1D and 2D masks had 
the same aperture size $d=25\ \mu m$ and the spacing step of $2d$. }

\label{1D_mask_design}
\end{figure}

Both 1D and 2D masks were manufactured by laser milling on a steel sheet with a thickness 
of a few micrometers. The apertures in the masks had a width of
$d = 25\ \mu m$ and the smallest spacing between them of $2d$, which equals 50 $\mu m$
(see Fig. \ref{1D_mask_design}). The single-shot diffraction patterns in horizontal 
and vertical direction for the 1D mask, as well as the diffraction pattern obtained 
from the 2D mask, were used to retrieve both the spatial coherence function and 
the ratios of intensities on the apertures employing the retrieval procedure 
described in "Methods".

\begin{figure}[H]
\centering
\begin{multicols}{2}
	\begin{subfigure}{0.5\textwidth}
	\centering  
    \includegraphics[scale=0.7]
    {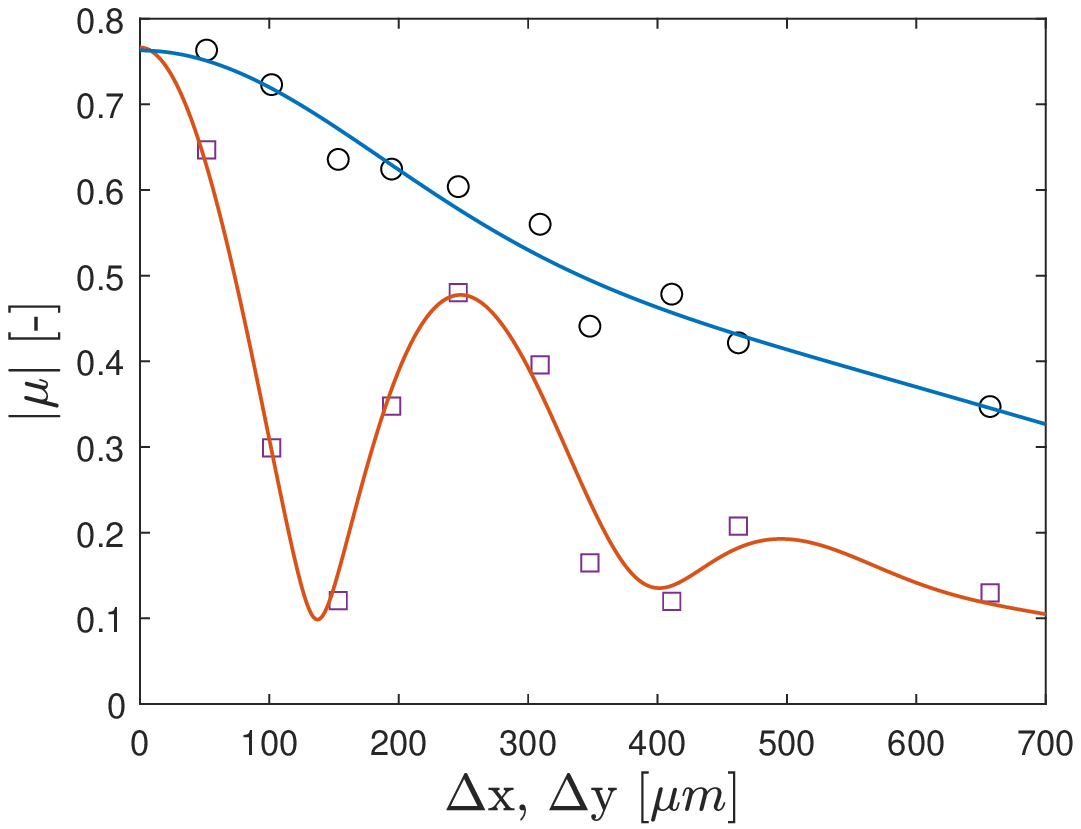}  
    \caption{ } % subcaption
	\end{subfigure} 
    \par
    \begin{subfigure}{0.5\textwidth}
    \centering   
    \includegraphics[scale=0.7]
    {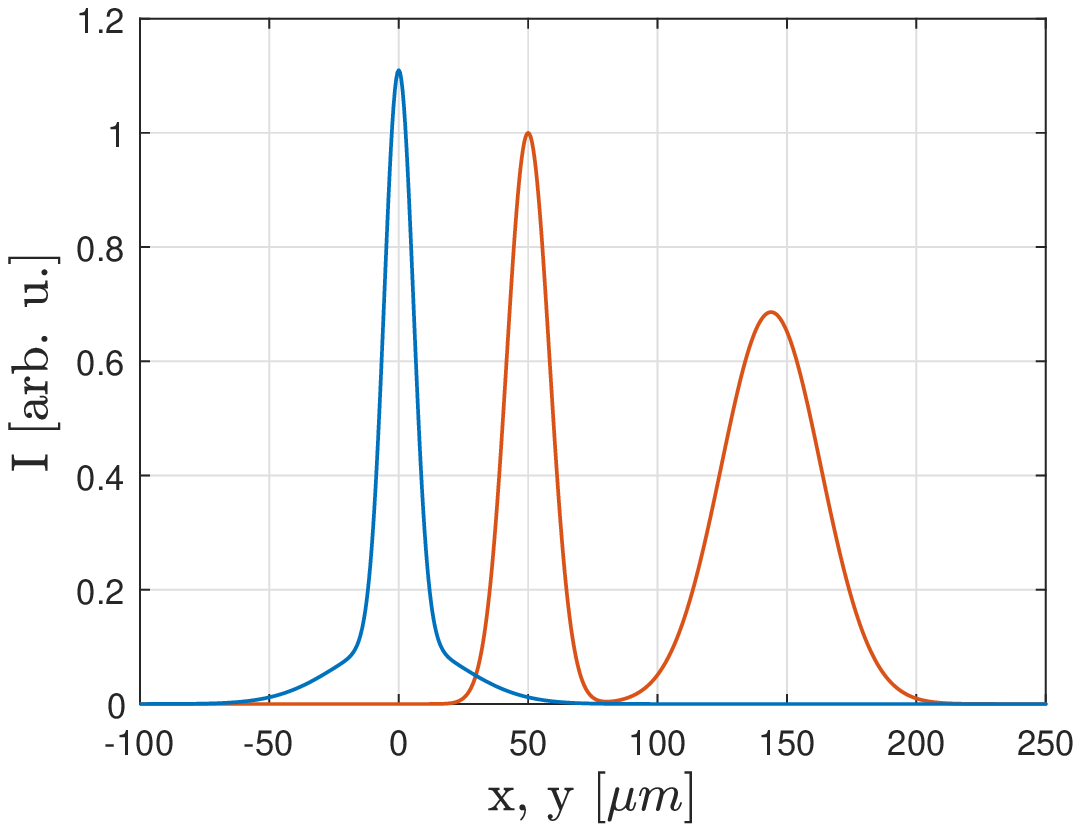}
    \caption{ } % subcaption
	\end{subfigure} 
    \par  
\end{multicols}
\caption{
(a): Modulus of degree of spatial coherence $|\mu|$ at the mask plane of Ne-like Zn SXRL 
retrieved from diffraction patterns of the measurements performed in horizontal (black circles) 
and vertical (purple squares) orientation of the 1D mask fitted by theoretical source 
models which were used as input for far field van Cittert-Zernike theorem. 
(b): 1D intensity distributions at the source plane of Ne-like Zn SXRL in vertical (red 
line, plotted with offset of 50 $\mu m$) and horizontal (blue line) orientation corresponding 
to the fitted model source.}

\label{1D_coherence_degree}
\end{figure}

\begin{figure}[H]
\centering
\begin{multicols}{2}
	\begin{subfigure}{0.5\textwidth}
	\centering  
    \includegraphics[scale=0.5]
    {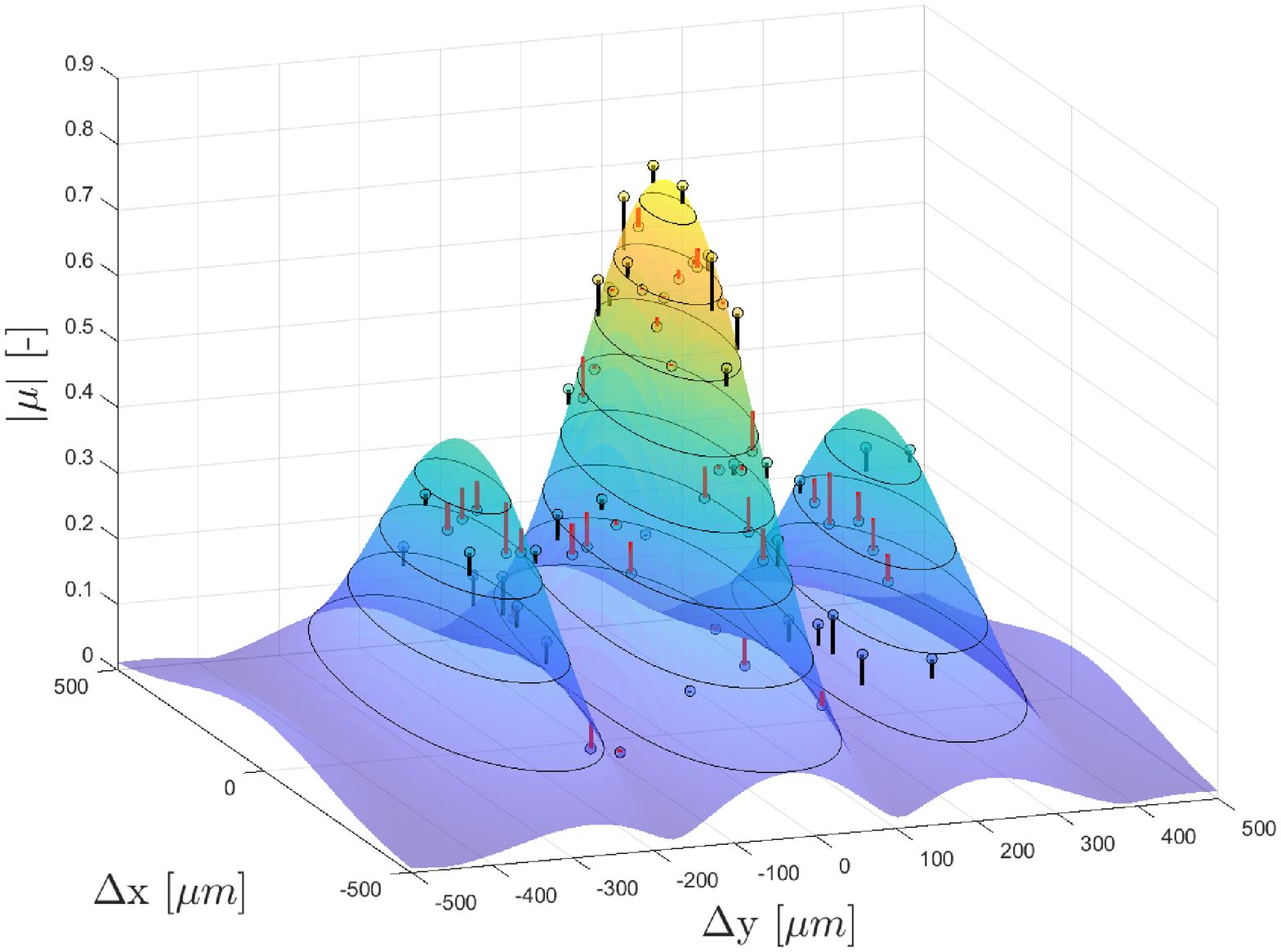}  
    \caption{ } % subcaption
	\end{subfigure} 
    \par
    \begin{subfigure}{0.5\textwidth}
    \centering   
    \includegraphics[scale=0.5]
    {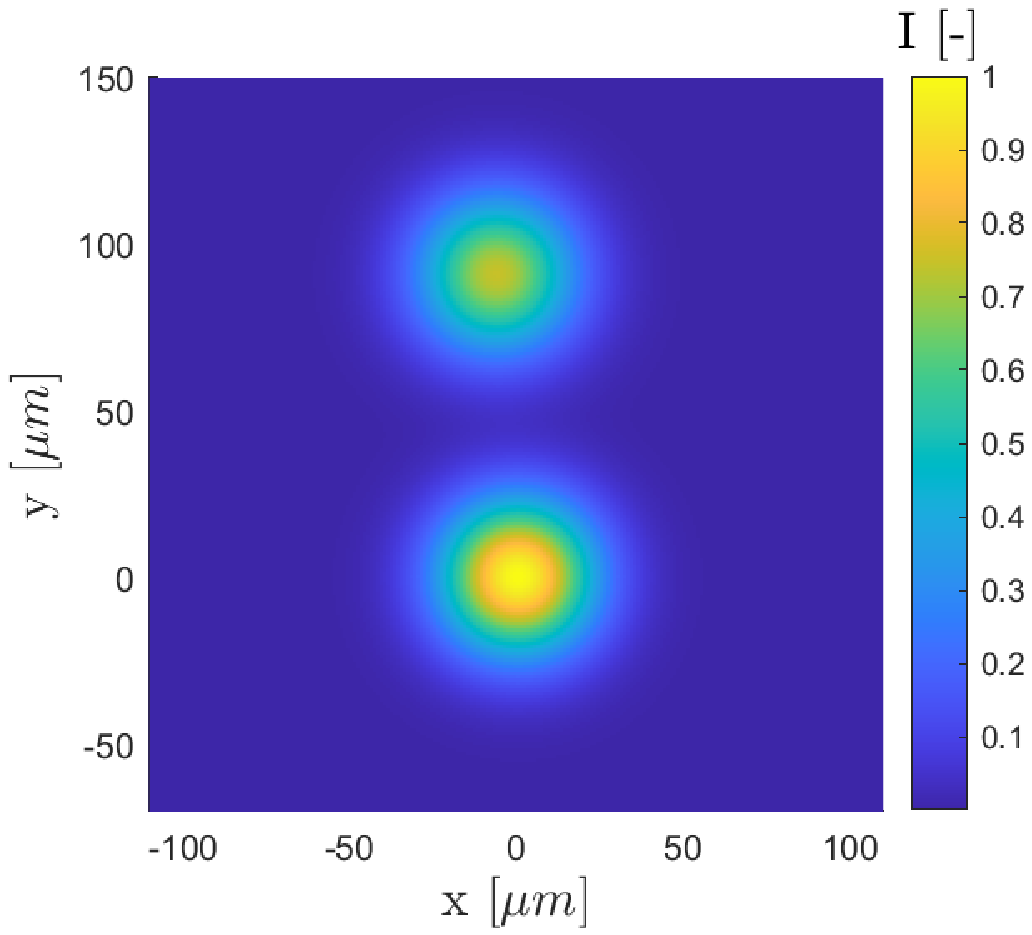}
    \caption{ } % subcaption
	\end{subfigure} 
    \par  
\end{multicols}
\caption{
(a): Modulus of degree of spatial coherence at the mask plane
 retrieved from the diffraction patterns of 2D measurement (black circles) and 
 fitted by theoretical source model used as input for far field van Cittert-Zernike 
 theorem (surface plot). Red and black lines show the differences between the measured 
 values and the fit. (b): 2D intensity distribution at the source plane of Ne-like 
 Zn SXRL corresponding to the fitted model source. }

\label{2D_coherence_degree}
\end{figure}

The Fourier transforms of the diffraction patterns which resemble the autocorrelation 
function of the masks are shown in Fig. \ref{1D_mask_design} together with the intensity 
distribution on the apertures (Fig. \ref{1D_mask_design}c and \ref{1D_mask_design}d) for 
each measurement. The degree of spatial coherence, retrieved from the recorded diffraction 
patterns, can be found in Fig. \ref{1D_coherence_degree} and Fig. \ref{2D_coherence_degree} 
in the case of 1D and 2D mask, respectively. 
It is worth noting that the intensity distribution on the apertures is an inseparable part of 
each measurement result, as no prior assumptions are made regarding the beam intensity.

It is evident that the spatial coherence function of the beam exhibits substantial 
differences between the vertical and horizontal directions. 
This behavior is due to the asymmetry of the source,
which has been previously observed directly \cite{j_polan}. The source asymmetry 
may arise from inhomogeneous pumping of the lasing medium or strong transverse 
plasma density gradients, affecting the propagation of soft X-ray radiation 
within the plasma.

To get more insight into the results, the resulting degree of spatial coherence 
for each measurement (corresponding to each laser shot) is fitted by a simple 
model assuming a coherence function of an uncorrelated radiation source 
consisting of two spatially separated sources
with symmetrical Gaussian intensity profiles. In this case, the degree of coherence 
is calculated by employing the van Cittert-Zernike theorem \cite{mandel}. 
The parameters characterizing the fitted source model are listed in Tab. \ref{table_two_gauss}
for each measurement. Although the number of data points of spatial coherence function 
(10 points for the 1D measurement and 36 points for the 2D measurement) could allow for a 
more complex model with additional degrees of freedom, we believe that our model 
(with 4 degrees of freedom in 1D case and 5 degrees of freedom in the 2D 
case) adequately represents the coherence properties of the source. This is 
supported by very good agreement between the fitted model and measured coherence function data,
as depicted in Figs. 4 and 5. Moreover, when 1D and 2D experimental results are 
compared, the spatial separation of the two sources in the vertical direction and the
obtained parameters of the fitting model are illustrating similar behavior. 
The main differences lie in the size and relative intensity of the 
two Gaussian sources. These differences can be attributed to the substantial 
shot-to-shot fluctuations characteristic of the soft X-ray laser.

One can notice that the measured degree of the spatial coherence shown in 
Fig. \ref{1D_coherence_degree} and Fig. \ref{2D_coherence_degree} does not 
asymptotically approach unity at zero separation of apertures. 
We believe this is due to incoherent radiation from the plasma that is 
still reflected by the multilayer mirror and various sources of detection noise.

\begin{table}[H]
\centering
% \hspace{-0.8cm}
\renewcommand{\arraystretch}{2.2}
\begin{tabular}[c]{|*{7}{c|}}
\hline 
    \multicolumn{2}{|c|}{ \textbf{Theoretical source model} } &
	$\bm{\frac{B}{A} [-]} $ &
	$\bm{ w_A [\mu m]} $ &
	$\bm{ w_B [\mu m]} $  &
	$\bm{ x_0 [\mu m]} $ &
	$\bm{ y_0 [\mu m]} $ 
	\\
\hline
\hline \textbf{1D (hor.):} & $Ae^{-2\frac{ x^2 }{w_A^2}} + 
							Be^{-2\frac{ (x-x_0)^2 }{w_B^2}}$ & 
					0.11 & 11 & 24 & 0 & - \\
\hline \textbf{1D (vert.):} & $Ae^{-2\frac{ y^2 }{w_A^2}} + 
							Be^{-2\frac{ (y-y_0)^2 }{w_B^2}}$ & 
					0.67 & 17 & 39 & - & 94 \\
\hline \textbf{2D:} & $Ae^{ -2\frac{ x^2 + y^2 }{w_A^2}} + 
							Be^{-2\frac{(x-x_0)^2 + (y-y_0)^2}{w_B^2}}$ & 		
					0.74 & 33 & 34 & -6 & 91 \\
\hline
\end{tabular}
\caption{ Parameters of the model source consisting of two spatially separated 
gaussian sources found by fitting the degree of coherence retrieved from the 
experimental data.}
\label{table_two_gauss}
\end{table}

\section*{Summary}

In conclusion, we have proposed and experimentally verified a new method for measuring 
the spatial coherence in a single-shot, without relying on any prior assumptions
about coherence or intensity distribution. We have demonstrated this method by measuring 
the 1D and 2D degree of spatial coherence of the beam generated from a plasma-based 
soft X-ray laser. The significant asymmetry observed in the spatial coherence 
function of the beam suggests that the commonly used Gauss-Schell beam model is 
not a sufficiently accurate description. We believe that our method represents a 
paradigm shift in the measurement of the spatial coherence of X-ray radiation, 
in particular.

\noindent \textbf{Funding:} This work was funded from ADONIS (CZ.02.1.01/0.0/0.0/16{\_}019/0000789)
    and LM2023068 Ministry of Education, Youth and Sports project supporting open access 
    to Large Research Infrastrutures.
\\
\textbf{Acknowledgement:} We are grateful to J. Hřebíček, T. Medřík, and J. Golasowski
for their support during experimental runs at PALS. % and to ? for manuscript revision. 

\section*{Methods}

\subsection*{NRA mask design}
There are several methods for designing NRAs, such as the method of minimum moment 
of inertia \cite{golay}. In our case, the masks were designed on a uniform discrete 
grid (orthogonal in the case of two dimensional mask) with spacing of $2d$, 
where $d$ is the size of the aperture. After placing the first aperture at 
the origin, an algorithm added apertures on the grid points 
only if the newly formed aperture array was nonredundant, 
while gradually increasing the distance from the origin.

\subsection*{Coherence retrieval algorithm}
Our method is based on retrieving the coherence function iteratively from a far-field 
diffraction pattern of non-redundant array (NRA) of apertures. By employing the modal 
expansion of the transverse coherence function $\Gamma(\bm{x_1}, \bm{x_2}, z)$, 
where $x_1$ and $x_2$ are vectors in the plane parallel to the field 
propagation direction $z$ reads

\begin{equation}
\Gamma( \bm{x_1}, \bm{x_2}, z) = \sum_n \lambda_n \phi_n^*(\bm{x_1}, z ) \phi_n( \bm{x_2}, z)
							   = \sum_n f_n^*(\bm{x_1}, z ) f_n( \bm{x_2}, z) \quad,				 
\label{coherence_mode_expansion}
\end{equation}
where $f_n(\bm{x}, z ) = \sqrt{\lambda_n}\phi_n(\bm{x}, z ) $ and $\phi_n(\bm{x}, z )$ are the
solutions of the integral equation \\
$ \int_{-\infty}^{+\infty} \Gamma(\bm{x_1}, \bm{x_2},z)\phi_n(\bm{x_1},z) \ \mathrm{d}\bm{x_1} =
\lambda_n \phi_n(\bm{x_2},z)   $ \cite{mandel}.

In other words, the light in certain plane can be described as an incoherent sum of 
coherent fields (modes), so knowing these fields in one plane, we can solve their 
propagation to another plane by well-known scalar field methods using e.g. Fresnel 
approximation of the diffraction integral \cite{goodman_1996}. Therefore, the 
relation \eqref{coherence_mode_expansion} allows 
us to propagate the spatial coherence function by independent propagation of multiple 
coherent modes $f_n$.

In our situation, where the intensity profile is unknown, we have employed an iterative 
algorithm for the spatial coherence function reconstruction \cite{rydberg}. The algorithm 
was modified to combine information from the measured 
far-field diffraction pattern and the knowledge of the aperture mask. This modification 
is similar to the approach used in phase retrieval algorithms employed in coherent 
diffractive imaging \cite{fienup, fienup2, gerchberg}, with the difference in 
reconstruction of multiple independent coherent modes.

\begin{figure}[H]

\centering  
\includegraphics[scale=0.8]
  {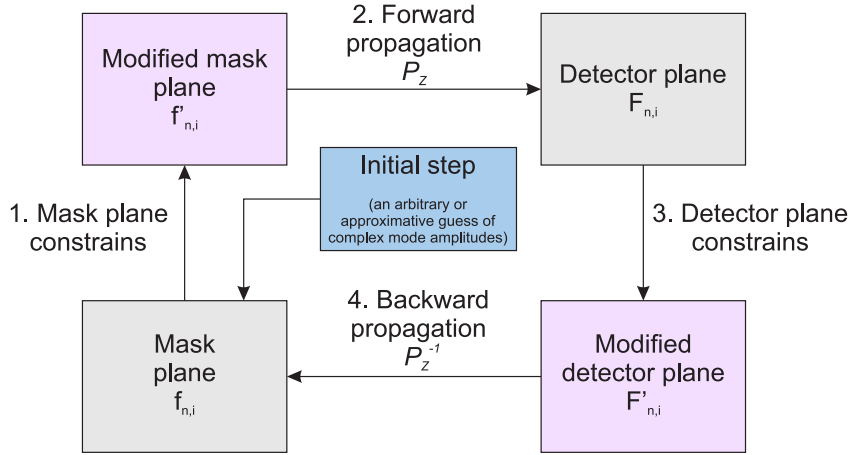}  
\caption{Block diagram of the algorithm for retrieval of coherence function from 
the diffraction on a nonredundant array of apertures.}
\label{coherence_block_diagram}

\end{figure}

By using a binary mask with NRA, which imposes a strong constraint on 
the field in the mask plane, and measuring its far-field diffraction pattern 
that forms the second constraint, we are able to retrieve a consistent set of coherent fields. 
These mutually incoherent fields allow us to obtain, in a single measurement, the coherence 
function, subsequently the degree of spatial coherence and the intensities on apertures 
according to following relations
\begin{align}
\mu(\bm{x_1},\bm{x_2}) &= \frac{\Gamma(\bm{x_1},\bm{x_2})}{\sqrt{I(\bm{x_1})I(\bm{x_2})}} \quad \\
I(\bm{x_i}) &= \Gamma(\bm{x_i}, \bm{x_i})	\quad ,
\end{align}

where $\Gamma(\bm{x_1},\bm{x_2})$ is obtained by the retrieved fields \eqref{coherence_mode_expansion}. 
The algorithm for this technique consists of following steps 
(depicted in Fig. \ref{coherence_block_diagram}):

\begin{enumerate}
\item[] 	\textbf{Initialization:} A suitable base of independent coherent modes 
				$\{f_1, f_2, ..., f_N \}$ in the mask plane has to be chosen. This can be
				e.g. a set of Hermite-Gauss polynomials with a guess of complex mode
				amplitudes $\lambda_n$ in accordance with equation
				\eqref{coherence_mode_expansion}.

\item 	\textbf{Modified mask plane:} The mask plane 
	constrains need to be applied for each mode $f_i$. These constrains 
	(known binary mask of the NRA) can be described by a support 
	function $\Pi(\bm{x})$ which is unitary everywhere where the 
	mask is transparent, otherwise it is zero. The modified field
	of each base function reads 
		
	\begin{equation} 
	f_{n,i}'(\bm{x}, 0) = f_{n,i}(\bm{x}, 0)\Pi(\bm{x}) \quad,
	\end{equation} 
    
    where $n$ and $i$ denotes the basis function and the number of iteration,
    respectively.

\item \textbf{Forward propagation:} Each modified base function is then propagated with 
	corresponding scalar method to detector plane. The propagation method is 
	represented by the propagation operator $\mathcal{P}_z$ as

	\begin{equation} 
	f_{n,i}(\bm{x},z) = \mathcal{P}_z\{f'_{n,i}(\bm{x},0)\} \quad.
	\end{equation} 
	
	Further, the calculated intensity on the detector plane in distance $z$ from
	the mask is obtained by incoherent sum of all modes
	 
	\begin{equation}
	I_i(\bm{x},z) = \sum_n |f_{n,i}(\bm{x},z)|^2 \quad.
	\label{mode_intensity}
	\end{equation}

\item \textbf{Detector plane constraints:} 

	Let $I_{ref}(\bm{x},z)$ denote the recorded diffraction pattern 
	intensity. The retrieved intensity from \eqref{mode_intensity} 
	has to be equal to the measured one, therefore the amplitude of each mode
	across the plane has to be normalized according to the relation 
	
	\begin{equation} 
	f_{n,i}'(\bm{x},z) = \sqrt{\frac{I_{ref}(\bm{x},z)}{I_i(\bm{x},z)}}
		f_{n,i}(\bm{x},z) \quad,
	\end{equation}
	
	while the phase of each field is preserved.

\item \textbf{Backward propagation:}
	Inversely to the forward propagation, each modified mode function $f_{n,i}'(\bm{x},z)$
	is propagated to the mask plane by the inverse propagator $\mathcal{P}_z^{-1}$  

	\begin{equation} 
	f_{n,i+1}(\bm{x},0) = \mathcal{P}_z^{-1}\{f'_{n,i}(\bm{x},z)\} \quad.
	\end{equation}
			
\end{enumerate}

The algorithm repeats the cycle from step 1 through step 4 in parallel for all independent
fields until the calculated far-field diffraction pattern $I_i(x, z)$ matches the measured 
pattern $I_{ref}(\bm{x},z)$ with sufficient precision. A consistent result of the retrieval 
algorithm can be indicated by reduction of the error 
residual function below a certain value, analogically to standard phase retrieval 
algorithms. It should be noted that the condition of oversampling of the measured 
diffraction pattern needs to be fulfilled, in order to achieve successful convergence. 

\bibliography{references_ssc}

\begin{thebibliography}{10}

\bibitem{mandel}
L.~Mandel and E.~Wolf, ``Optical coherence and quantum optics,'' {\em Cambridge
  University Press, New York USA}, 1995.

\bibitem{gartside_2010}
L.~M.~R. Gartside, G.~J. Tallents, A.~K. Rossall, E.~Wagenaars, D.~S.
  Whittaker, M.~Kozlová, J.~Nejdl, M.~Sawicka, J.~Polan, M.~Kalal, and B.~Rus,
  ``Extreme ultraviolet interferometry of warm dense matter in laser plasmas,''
  {\em Optics Letters}, vol.~35, no.~22, p.~3820–3822, 2010.

\bibitem{young_1804}
T.~Young, ``Experimental demonstration of the general law of the interference
  of light,'' {\em Philosophical Transactions of the Royal society of London},
  vol.~94, p.~1–16, 1804.

\bibitem{liu}
Y.~Liu, Y.~Wang, M.~A. Larotonda, B.~M. Luther, J.~J. Rocca, and D.~T. Attwood,
  ``Spatial coherence measurements of a 13.2 nm transient nickel-like cadmium
  soft x-ray laser pumped at grazing incidence,'' {\em Optics Express},
  vol.~14, no.~26, pp.~12872--12879, 2006.

\bibitem{vartanyants}
I.~A. Vartanyants, A.~Singer, A.~P. Mancuso, O.~M. Yefanov, A.~Sakdinawat,
  Y.~Liu, E.~Bang, G.~J. Williams, G.~Cadenazzi, B.~Abbey, H.~Sinn, D.~Attwood,
  K.~A. Nugent, E.~Weckert, T.~Wang, D.~Zhu, B.~Wu, C.~Graves, A.~Scherz, J.~J.
  Turner, W.~F. Schlotter, M.~Messerschmidt, J.~Lüning, Y.~Acremann,
  P.~Heimann, D.~C. Mancini, V.~Joshi, J.~Krzywinski, R.~Soufli,
  M.~Fernandez-Perea, S.~Hau-Riege, A.~G. Peele, Y.~Feng, O.~Krupin,
  S.~Moeller, and W.~Wurth, ``Coherence properties of individual femtosecond
  pulses of an x-ray free-electron laser,'' {\em Physical Review Letters},
  vol.~107, no.~144801, 2011.

\bibitem{singer_1}
A.~Singer, F.~Sorgenfrei, A.~P. Mancuso, N.~Gerasimova, O.~M. Yefanov,
  J.~Gulden, T.~Gorniak, T.~Senkbeil, A.~Sakdinawat, Y.~Liu, D.~Attwood,
  S.~Dziarzhytski, D.~D. Mai, R.~Treusch, E.~Weckert, T.~Salditt, A.~Rosenhahn,
  W.~Wurth, and I.~A. Vartanyants, ``Spatial and temporal coherence properties
  of single free-electron laser pulses,'' {\em Optics Express}, vol.~20,
  no.~16, p.~17480, 2012.

\bibitem{singer_2}
A.~Singer, I.~A. Vartanyants, M.~Kuhlmann, S.~Duesterer, R.~Treusch, and
  J.~Feldhaus, ``Transverse-coherence properties of the free-electron-laser
  flash and desy,'' {\em Physical Review Letters}, vol.~101, no.~254801, 2008.

\bibitem{nugent}
K.~A. Nugent and J.~E. Trebes, ``Coherence measurement technique for
  short-wavelength light sources,'' {\em Review of Scientific Instruments},
  vol.~63, p.~2146, 1992.

\bibitem{trebes}
J.~E. Trebes, K.~A. Nugent, S.~Mrowka, R.~A. London, T.~W. Barbee, M.~R.
  Carter, J.~A. Koch, B.~J. MacGowan, D.~L. Matthews, L.~B.~D. Silva, G.~F.
  Stone, and M.~D. Feit, ``Measurement of the spatial coherence of a soft-x-ray
  laser,'' {\em Physical Review Letters}, vol.~68, no.~5, p.~588, 1992.

\bibitem{gonzales}
A.~I. González and Y.~Mejía, ``Nonredundant array of apertures to measure the
  spatial coherence in two dimensions with only one interferogram,'' {\em J.
  Opt. Soc. Am. A}, vol.~28, no.~6, 2011.

\bibitem{mejia}
Y.~Mejía and A.~I. González, ``Measuring spatial coherence by using a mask
  with multiple apertures,'' {\em Optics Communications}, vol.~273,
  pp.~428--434, 2007.

\bibitem{j_duarte}
J.~Duarte, A.~I. González, R.~Cassin, R.~Nicolas, M.~Kholodstova, W.~Boutu,
  M.~Fajardo, and H.~Merdji, ``Single-shot spatial coherence characterization
  of x-ray ultrafast sources,'' {\em Optics Letters}, vol.~46, p.~7, 2021.

\bibitem{rydberg}
C.~Rydberg and J.~Bengtsson, ``Numerical algorithm for the retrieval of spatial
  coherence properties of partially coherent beams from transverse intensity
  measurements,'' {\em Optics Express}, vol.~15, no.~21, p.~13613, 2007.

\bibitem{b_rus}
B.~Rus, T.~Mocek, A.~R. Präg, M.~Kozlová, G.~Jamelot, A.~Carillon, D.~Ros,
  D.~Joyeux, and D.~Phalippou, ``Multimillijoule, highly coherent x-ray laser
  at 21 nm operating in deep saturation through double-pass amplification,''
  {\em Physical review A}, vol.~66, no.~063806, 2002.

\bibitem{j_polan}
J.~Polan, T.~Mocek, M.~Kozlová, P.~Homer, and B.~Rus, ``Spatial and temporal
  profiles of the 21.2-nm saturated x-ray laser output,'' {\em X-Ray Lasers
  2006, Springer Proceedings in Physics}, vol.~115, pp.~139--147, 2006.

\bibitem{golay}
M.~J.~E. Golay, ``Point arrays having compact, nonredundat autocorrelations,''
  {\em J. Opt. Soc. Am.}, vol.~61, no.~2, pp.~272--273, 1971.

\bibitem{goodman_1996}
J.~W. Goodman, {\em Introduction to Fourier Optics}.
\newblock 1996.

\bibitem{fienup}
J.~R. Fienup, ``Phase retrieval algorithms: a comparison,'' {\em Applied
  Optics}, vol.~21, no.~15, p.~2758, 1982.

\bibitem{fienup2}
J.~R. Fienup, ``Reconstruction of an object from the modulus of its fourier
  transform,'' {\em Optics Letters}, vol.~3, no.~1, pp.~27--29, 1978.

\bibitem{gerchberg}
R.~W. Gerchberg and W.~O. Saxton, ``A practical algorithm for the determination
  of phase from image and diffraction plane pictures,'' {\em Optik}, vol.~35,
  no.~2, pp.~237--246, 1972.

\end{thebibliography}
\bibliographystyle{ieeetr}

\end{document}